\documentclass[seceq,preprint]{ptptex}
\title{New Parameterization in Muon  Decay\\ 
and the Type of Emitted Neutrino II%
\footnote{Published in Progress of Theoretical Physics~{\bf 118}, No.6~(2007), pp.1069-1086; \\ 
\hspace{0.9cm} see also,  Errata: {\it ibid}.~{\bf 122}, No.3~(2009), p.805.}}
\author{Masaru \textsc{Doi},$^{1,}$
\footnote{E-mail: doi@gly.oups.ac.jp} 
Tsuneyuki \textsc{Kotani}$^{2,}$
\footnote{E-mail: tsune.kotani@nifty.com} 
and Hiroyuki \textsc{Nishiura}$^{3,}$
\footnote{E-mail: nishiura@is.oit.ac.jp}}

\inst{$^1$Osaka University of Pharmaceutical Sciences, \\
Takatsuki, Osaka 569-1094, Japan\\
$^2$2-8-26, Higashitoyonaka,
Toyonaka, Osaka, 560-0003, Japan\\
$^3$Faculty of Information Science and Technology, \\
Osaka Institute of Technology, Hirakata, Osaka 573-0196, Japan.}

\abst{
In the previous paper, new sets of parameters in place of the Michel 
parameters have been proposed to analyze the data on the 
muon decay $\mu^{+} \to e^{+}\nu_{e}\overline{\nu_{\mu}}$.  Both 
$(V-A)$ and $(V+A)$ charged currents with the finite neutrino mass 
have been used.  In the present paper, this parameterization is 
extended to the more general form, and the method of data 
analysis (least squares) is discussed to determine the rate of contribution 
from the $(V+A)$ current.  There is a simple form in which a set of 
parameters is related primitively with the physical quantities.  
It is shown that the Michel parameters are one of the other 
sets which are obtained from this simple form by rearranging one term 
in it.  We derive the condition to get the equivalent information 
on the unknown physical quantities, when the data are analyzed 
by using these simple and rearranged forms separately.  There is some 
possibility 
to get different results from these analyses, because the 
equivalent condition is very delicate and the QED radiative 
corrections should be treated carefully.  We propose a consistent 
formula for the data analysis.  It is useful to compare the value 
of the least squares for the simple form with the one for the 
prediction of the standard model, because the large difference 
is not expected, especially for the Majorana neutrino case.  
Although we proposed a method to discriminate the type of 
neutrino in the previous paper, 
it is shown that it is not correct.}

\begin{document}
\maketitle

%%%%%%%%%%%%%%%%%%%%%%%%%%%%%%%%%%%%%%%%%%%%%%%%%%%%%%%%%%%%%%%
%%%%%%%%%%%%%%%%%          Section 1         %%%%%%%%%%%%%%%%%%
%%%%%%%%%%%%%%%%%%%%%%%%%%%%%%%%%%%%%%%%%%%%%%%%%%%%%%%%%%%%%%%
\section{Introduction}

The normal muon decay 
%$\mu^{+} \to e^{+}\nu_{e}\overline{\nu_{\mu}}$ 
has been studied as a tool with high statistics to determine the 
structure of the weak interaction.  The purpose of this paper 
is to investigate the effect of the $(V + A)$ current added to 
the standard model and to find a means to treat the QED radiative 
corrections in the data analysis 
based on the method of least squares.  
Both the Dirac and Majorana neutrino cases are examined.

Recently, the TWIST group~\cite{Twist} has reported their precise 
experimental data and analyzed them by using the helicity 
preserving four fermion weak interaction with 
$(S \pm P),\,(V \pm A)$ and $T$ forms~\cite{Fetscher}. 
Their expression based on the Michel 
parameters for the $e^{+}$ energy spectrum is as follows; 
\begin{eqnarray}
\frac{d^2 \Gamma}{d x \, d \cos \theta}
     \propto
     \Bigl[ \,{\cal N}(x) 
      + P_{\mu} \, \cos \theta \, {\cal P}(x) \, \Bigr] ,
     \label{eq:1no01}
\end{eqnarray}
where 
\begin{eqnarray}
{\cal N}(x) &=& 6 \,x^{2} \, \Bigl[ (1 - x) +
     \frac{2}{9} \, (4 x - 3) \, \rho_{M} \Bigr] ,
     \label{eq:1no02} \\
{\cal P}(x) &=& 2 \, x^2 \, \xi_{M} \, 
     \Bigl[ (1 - x) + \frac{2}{3} \, ( 4 x - 3 ) \, \delta_{M}
     \Bigr] .
     \label{eq:1no03}
\end{eqnarray}
Here $x$ is defined as $ x  =  E/W$, where $E$ is the energy of 
the emitted positron 
and $W  = (m_{\mu}^{2} + m_{e}^{2} \,)/(2 \,  m_{\mu}) \,$, and   
$m_\mu$ and $m_e$ are the muon and positron masses, respectively.  
The angle $\theta$ is the direction of emitted $e^{+}$ 
with respect to the muon polarization vector 
$\vec{P_{\mu}}$ at the instant of $\mu^{+}$ decay.  
In the above expressions, we do not include terms proportional to $m_e$ 
and neutrino masses, and also QED radiative corrections in order 
to simplify our explanation. 

The standard model predicts $\rho_{M} = \delta_{M} = 0.75$ and 
$\xi_{M} = 1$ for these Michel parameters.  The traditional way 
of experimental data 
analysis has been to determine the deviations from these predicted 
values.  The new experimental center values reported by the TWIST 
group~\cite{Twist} are $\rho_{M} = 0.75080$, $\delta_{M}= 0.74964$ 
and $0.9960 <P_\mu \xi_{M}\leq \xi_{M}<1.0040$.  The QED radiative 
corrections are taken into account in their analysis.  As you see, 
these deviations are small, so that it is preferable to determine 
them directly.

In our previous paper~\cite{Doi2005} which is referred to as the 
paper I hereafter, we proposed to use new 
parameters that is suitable for investigating these 
deviations.  We showed that various parameterizations 
are allowed by adopting the possible choices of the normalization 
factor for the isotropic part of  energy spectrum. 
Among them, we mainly discussed the specific one 
which is related with the Michel parameters directly.  
This will be referred to as the Michel parameterization.  
We assumed that the weak interaction Hamiltonian consists of both 
$(V-A)$ and $(V+A)$ charged currents and that the neutrino 
has the finite mass.  
Therefore, we have two kinds of lepton mixing matrices and 
three weak coupling constants which represent the rate of mixture 
of $(V+A)$ current.  These unknown quantities will
be referred to as "the weak coupling constants" in short.

In this paper, the more general parameterization is investigated 
within the frame of the same Hamiltonian.  
In the simple form, a set of parameters is expressed primitively 
in terms of the combination of the weak coupling constants.  Meanwhile, 
it is shown that the specific form related with the Michel 
parameterization is one of many forms which are derived 
from the above simple form by rearranging a term in it.  
Sets of parameters in these rearranged forms 
are related with the combinations of the weak coupling constants 
in somewhat complicated manners.  
Of course, these different sets of parameters 
should offer the same information 
on the weak coupling constants.  In \S 3, we investigate the 
condition to get the same information mentioned above, when 
the experimental data are analyzed by using the method of 
the least squares for these different sets individually.  It is 
pointed out that this condition is very delicate and we should be 
careful in treating the QED radiative correction in the data analysis.

In the paper I, we also proposed a method of discriminating between 
the Dirac and Majorana neutrino experimentally by using the method of 
least squares for the $e^+$ energy spectrum.  However, we shall show in 
\S 2 of the present paper that this proposal is incorrect.  This 
discrimination is not easy in the muon decay.  It will be discussed 
in \S 4.

In \S 2, we present the comprehensive discussions on the 
general form of parameterization.  
The present experimental limits are listed for new parameters.  
In \S 3, we discuss the condition to get the information on 
the unknown weak coupling constants 
and the method of taking the QED radiative corrections 
into account appropriately.  We propose the consistent 
formula for the method of the least squares in the data analysis.  
In \S 4, we present some comments.  Appendix A contains expressions 
for the polarization of the emitted positron.

%%%%%%%%%%%%%%%%%%%%%%%%%%%%%%%%%%%%%%%%%%%%%%%%%%%%%%%%%%%%%%%
%%%%%%%%%%%%%%%%%          Section 2         %%%%%%%%%%%%%%%%%%
%%%%%%%%%%%%%%%%%%%%%%%%%%%%%%%%%%%%%%%%%%%%%%%%%%%%%%%%%%%%%%%
\section{\label{sec:two}Parameterization of the decay spectrum}

We assume the following form of effective weak interaction Hamiltonian 
for the $\mu^{+}$ decay,~\cite{Doi1} 
\begin{equation}
{\cal{H}}_W(x)=\frac{G_F}{\sqrt{2}} \left\{ 
    j_{eL\, \alpha}^\dagger j_{\mu L}^\alpha
    + \lambda j_{eR\,  \alpha}^\dagger j_{\mu R}^\alpha
    + \eta j_{eR\,  \alpha}^\dagger j_{\mu L}^\alpha
    +\kappa j_{eL\,  \alpha}^\dagger j_{\mu R}^\alpha
    \right\} +\mbox{H.c.}\ ,
    \label{eq:2no01}
\end{equation}
where $G_{F}$ is the Fermi coupling constant.~\cite{Ritbergen}  
Weak coupling constants ($\lambda, \, \eta$ and $\kappa$) 
represent the rate of mixture of the $(V + A)$ current for the 
combination of the left(right)-handed charged leptonic 
currents $j_{\ell L(R)}$.%%
%%%%%%%%%%%%% Footnote begins %%%%%%%%%%%%%%%%%
\footnote{In order to see the physical meaning of weak
coupling constants, let us consider a typical example of the 
gauge theory, that is, the $SU(2)_{L} \times SU(2)_{R} 
\times U(1)$ model with left- and right-handed weak gauge 
bosons, $W_{L}$ and $W_{R}$.  Weak coupling constants 
in Eq.~(\ref{eq:2no01}) are 
related to the physical quantities:
\begin{equation*}
\lambda  \sim  (\lambda_{c} + \tan^{2} \zeta),
\hspace{10mm}
\kappa  =  \eta \sim (- \tan \zeta).
\end{equation*}
Here $\lambda_{c} = (M_{1}/M_{2})^2$, where $M_{1}$ and $M_{2}$ 
are masses of the mass-eigenstate gauge bosons 
which are expressed in terms of the weak eigenstate gauge bosons 
$W_{L}$ and $W_{R}$ with the mixing angle $\zeta$. 
For example, see Appendix A of the paper I.}
%%%%%%%%%%%  Footnote end  %%%%%%%%%%%%%%%%%%%%%%%%
These currents are defined as
\begin{eqnarray}
     j_{\ell L\, \alpha}(x)
      &=& \sum_{j=1}^{2n}\overline{{E_{\ell}}(x)}\gamma_\alpha(1-\gamma_5)
      U_{\ell j}N_j(x),
      \label{eq:2no02-1}\\
     j_{\ell R\, \alpha}(x)
     &=& \sum_{j=1}^{2n}\overline{{E_{\ell}}(x)}\gamma_\alpha(1+\gamma_5)
     V_{\ell j}N_j(x),
     \label{eq:2no02-2}
\end{eqnarray} 
for the case of the $n$ generations. Here 
$U_{\ell j}$ and $V_{\ell j}$ are the left- and 
right- handed lepton mixing matrices, 
and $E_{\ell}$ and $N_j$ represent, respectively, 
the mass eigenstates of charged leptons and neutrinos.  
Throughout this paper, 
neutrinos are assumed to have finite masses.

The decay spectrum of $e^{+}$ in the rest frame of polarized 
$\mu^{+}$ is defined as
\begin{equation}
\frac{d^2 \Gamma}
     {d x \, d \cos \theta}
     =\Gamma_{W} \, A \,
     \bigl[ {\cal N}(x) + P_{\mu} \, \cos \theta \, {\cal P}(x) \bigr],
     \label{eq:2no02}
\end{equation}
where the sum over the spin of $e^{+}$ has been taken and
\begin{eqnarray}
\Gamma_{W}   &=&   \frac{m_{\mu} \, G_{F}^2 \, W^4}
     {12 \, \pi^3}\, .
       \label{eq:2no03}
\end{eqnarray}

The isotropic and anisotropic parts of energy spectrum obtained 
from the leptonic Hamiltonian in Eq.~(\ref{eq:2no01}) 
are expressed as follows:~\cite{Doi2, Doi2005} 
\begin{eqnarray}
{\cal N}(x)
  &=& \left( \frac{1}{A} \right)
  \left[ a_{+} \, n_{1}(x)
     +  ( \, k_{+ \, c} + \varepsilon_{m} \, k_{+ \, m} \, ) \, n_{2}(x) 
     + \varepsilon_{m} \, \lambda \, d_{r} \, n_{3}(x) \right],
     \label{eq:2no04} \\
{\cal P}(x)
 &=& \left( \frac{1}{A} \right) 
     \left[ a_{-} \, p_{1}(x)    
     + ( \, k_{- \, c} + \varepsilon_{m} k_{- \, m} \, ) \, p_{2}(x) \right],
     \label{eq:2no05}
\end{eqnarray}
where the decay formulae for the Dirac and Majorana neutrinos are 
obtained by setting $\varepsilon_{m} = 0$ 
and $\varepsilon_{m} = 1$, respectively. 

The constant $A$ is introduced in Eqs.~(\ref{eq:2no02}),
(\ref{eq:2no04}) and (\ref{eq:2no05}) to simplify the coefficient 
of the prediction by the standard model in ${\cal N}(x)$.  Its 
explicit form will be given later, see Eq.~(\ref{eq:2no17}).  
This $A$ will be called a normalization factor 
according to the paper I.

The $x$-dependent parts are defined as follows:
\begin{eqnarray}
n_{1}(x) &=& x_p \, (3 x - 2 x^{2} - x_{0}^{2}),
     \label{eq:2no11} \\
n_{2}(x) &=& 12 \, x_p \, x \, (1 - x),
     \label{eq:2no12} \\
n_{3}(x) &=& 6 \, x_p \, x_{0} \, (1-x),
     \label{eq:2no13} \\
p_{1}(x) &=& x_p^2 \,(-1 + 2 \, x - r_{0}^{2}),
     \label{eq:2no14} \\
p_{2}(x) &=& 12 \, x_p^2 \, (1 - x),
     \label{eq:2no15}
\end{eqnarray}
where
\begin{eqnarray}
x_{p} = \sqrt{x^2 - x_{0}^2}, \hspace{3mm}
     x_{0} = \frac{m_e}{W}=9.7 \cdot 10^{-3}, \hspace{3mm}
     r_{0}^2 = \frac{m_e^2}{m_{\mu} \, W}=4.7 \cdot 10^{-5}.
     \label{eq:2no16}
\end{eqnarray}

The first terms $n_{1}(x)$ and $p_{1}(x)$ in 
${\cal N}(x)$ and ${\cal P}(x)$, respectively, are predictions 
from the standard model.  They are named the standard functions 
 in this paper.  Meanwhile, the others, 
$n_{2}(x)$, $n_{3}(x)$ and $p_{2}(x)$, are called the 
deviation functions.  In these functions, all terms 
proportional to the neutrino mass are neglected, because of the 
smallness of $(m_{\nu}/m_{\mu}) < 9 \cdot 10^{-9}$.  Here, 
$m_{\nu}$ stands for a typical mass scale of emitted neutrinos 
and is taken to be $m_{\nu} < 1 \mathrm{eV}$.%%%%%  
%%%%%%%%%%%%%%%% Footnote begin ********************************
\footnote{\label{conditionA} 
The spectrum functions ${\cal N}(x)$ and ${\cal P}(x)$ in 
Eqs.~(\ref{eq:2no04}) and (\ref{eq:2no05}) 
are precise except in the very tiny range of order 
$(m_{\nu}/m_{\mu})^{2} < \mathrm{O(10^{-16})}$ from the maximum of 
$x$, $x_{\mathrm{max}}$, which is expressed as 
$x_{\mathrm{max}}=1-[(m_{j} + m_{k})^{2}/(2 \, m_{\mu} \, W)]$ 
by using masses of two emitted neutrinos ($m_{j}$ and $m_{k}$).  
This is because two kind of kinematical factors come from the phase 
space integral over emitted neutrinos and they 
can be set unity in the almost entire range of $x$.  
In other words, they show significant $x$-dependence only 
in the extremely narrow range of order 
$(m_{\nu}/m_{\mu})^{2}$ near $x_{\mathrm{max}}$.  
It should be noticed that ${\cal N}(x)$ and ${\cal P}(x)$ become 
zero sharply at $x_{\mathrm{max}}$ due to these kinematical factors.  
For details, see \S 2.1 of the paper I.}
%%%%%%%%%%%%%%% Footnote end ************************************

Coefficients $a_{\pm}$, $k_{\pm \, c}$, $k_{\pm \, m}$ and 
$d_{r}$ in ${\cal N}(x)$ and ${\cal P}(x)$ are constants.  
They are obtained by summing up some products of 
the weak coupling constants ($\lambda, \, \eta$ and $\kappa$) 
and the lepton mixing matrices ($U_{\ell \, j}$ and $V_{\ell \, j}$) 
only over the emitted neutrinos.  
In the Dirac neutrino case, they are defined by
\begin{eqnarray}
a_{\pm} = \left( 1 \pm \lambda^2 \right),
\hspace{10mm}
k_{\pm \, c} = \left( \frac{1}{2} \right)
     \left( \kappa^2 \pm \eta^{2} \right),
     \label{eq:2no06}
\end{eqnarray}
Here it is assumed that all neutrinos can be emitted 
in the $\mu$ decay, and we have used the relation
\begin{eqnarray}
{\mathrm{\Sigma}}_{j}|U_{\ell j}|^2 = 
     {\mathrm{\Sigma}}_{j}|V_{\ell j}|^2 = 1,
     \label{eq:2no04-1}
\end{eqnarray} 
from the unitarity condition, because $j$ 
in the sum runs over all $n$ neutrinos.

By contrast, in the Majorana neutrino case, we assume 
that there exist additional $n$ heavy neutrinos 
which are not emitted in the decay.  
Then, these coefficients are given as follows:
\begin{eqnarray}
a_{\pm} &=&
     \left[ \left( 1 - \overline{u_{e}}^{\, 2} \right)
     \left( 1 - \overline{u_{\mu}}^{\, 2} \right)
     \pm \lambda^2 \, \overline{v_{e}}^{\, 2} \,
     \overline{v_{\mu}}^{\, 2} \right] ,
     \label{eq:2no07} \\
k_{\pm \, c} &=& \left( \frac{1}{2} \right)
     \left[ \kappa^2 \, (1 - \overline{u_{e}}^{\, 2} )
      \, \overline{v_{\mu}}^{\, 2}
      \pm \eta^{2} \, \overline{v_{e}}^{\, 2} \,
      (1 - \overline{u_{\mu}}^{\, 2} ) \right] ,
     \label{eq:2no08} \\
k_{\pm \, m} &=& \left( \frac{1}{2} \right)
     \left[ \,\kappa^2 \,
     | \, \overline{w_{e \mu}} \, |^{2}
     \pm \eta^{2} \,
     | \, \overline{w_{e \mu \, h}} \, |^{2} \, \right] \, ,
     \label{eq:2no09} \\
d_{r} &=& \left( \frac{1}{2} \right)
     \mbox{Re} ( \overline{w_{e \mu}}^{\, *} \,\,
     \overline{w_{e \mu \, h}} ).
     \label{eq:2no10}
\end{eqnarray}
Here, $\overline{u_{\ell}}^{\, 2}$, $\overline{v_{\ell}}^{\, 2}$, 
$\overline{w_{e \mu}}$ and $\overline{w_{e \mu h}}$ 
are all small quantities which stand for the extent of deviations 
from the unitarity condition due to the existence of heavy neutrinos.  
\begin{align}
    &{\mathrm{\Sigma}}_{j}^{ \, \prime}|U_{\ell j}|^2 
    \equiv 1-\overline{u_{\ell}}^{\, 2},
    & &
    {\mathrm{\Sigma}}_{j}^{ \, \prime}|V_{\ell j}|^2 
    \equiv \overline{v_{\ell}}^{\, 2},
    \label{eq:2no05-1} \\
    &{\mathrm{\Sigma}}_{j}^{\, \prime} \, U_{ej} \, V_{\mu j}
     \equiv \overline{w_{e \mu}},
    & &
    {\mathrm{\Sigma}}_{k}^{\, \prime} \, V_{ek} \, U_{\mu k}
    \equiv \overline{w_{e \mu \, h}},
    \label{eq:2no05-2} 
\end{align}
where the primed sum is taken over only 
$n$ light neutrinos out of $2n$ neutrinos.  
Their orders of magnitudes are 
$\overline{u_{\ell}}^{\, 2} \sim \overline{v_{\ell}}^{\, 2} 
\sim \mathrm{O}((m_{\nu D}/m_{\nu R})^2)$ and  
$\overline{w_{e \mu}} \sim \overline{w_{e \mu h}} 
\sim \mathrm{O}(m_{\nu D}/m_{\nu R})$, 
if the seesaw mechanism is assumed.%% 
%%%%%%%%%%%%% Footnote begins %%%%%%%%%%%%%%%%%
\footnote{For details, see Appendix A and \S 2.2 of the paper I 
as an example.}
%%%%%%%%%%%  Footnote end  %%%%%%%%%%%%%%%%%%%%%%%%
Here, $m_{\nu D}$ and $m_{\nu R}$ are, respectively, 
representatives of Dirac type and right-handed Majorana type 
masses in the neutrino mass matrix.

\subsection{Isotropic part of the spectrum: ${\cal N}(x)$}
Let us consider the isotropic part ${\cal N}(x)$.  
For the aim to survey the deviation from the standard model, 
it is suitable to examine ${\cal N}(x)$ by treating the standard 
function $n_{1}(x)$ as a base of analysis.  In 
order to see the relation with the Michel parameter introduced 
in Eq.~(\ref{eq:1no02}), we rearrange $n_{1}(x)$ in ${\cal N}(x)$ 
as follows:
\begin{eqnarray}
{\cal N}(x) 
     &=& \frac{1}{A}\Big\{n_{1}(x) [a_{+} +  s \, (k_{+ \, c}
     + \varepsilon_{m} \, k_{+ \, m} )
     + t \, \varepsilon_{m} \, \lambda \, d_{r}] 
     \nonumber\\
     &&+ [n_{2}(x)- s \, n_{1}(x) ] \, (k_{+ \, c}
     + \varepsilon_{m} \, k_{+ \, m} )
     + [n_{3}(x) - t \, n_{1}(x)] \, \varepsilon_{m} \, \lambda \, d_{r}\Big\},
     \label{eq:2no18a}
\end{eqnarray}
where $s$ and $t$ are some arbitrary numbers.  
Here the normalization factor $A$ is set to the following 
${\cal A}_{s \, t}$ to simplify the coefficient of $n_{1}(x)$;
\begin{eqnarray}
{\cal A}_{s \, t} = a_{+} +  s \, (k_{+ \, c}
     + \varepsilon_{m} \, k_{+ \, m} )
     + t \, \varepsilon_{m} \, \lambda \, d_{r} > 0.
     \label{eq:2no17}
\end{eqnarray}
It is natural to restrict $s$ and $t$ to the values satisfying 
the condition ${\cal A}_{s \, t} > 0$.

Thus, the isotropic part is denoted as 
${\cal N}_{s \, t}(x)$ which takes the following form:
\begin{equation}
{\cal N}_{s \, t}(x) 
     = n_{1}(x) + [n_{2}(x)- s \, n_{1}(x) ] \, \rho^{(s \, t)}
     + [n_{3}(x) - t \, n_{1}(x)] \, \eta^{(s \, t)} ,
     \label{eq:2no18}
\end{equation}
where two parameters $\rho^{(s \, t)}$ and $\eta^{(s \, t)}$ 
are defined as 
\begin{eqnarray}
\rho^{(s \, t)} = \frac{k_{+ \, c} + \varepsilon_{m} k_{+ \, m}}
     {{\cal A}_{s \, t}} > 0, \hspace{15mm}
     \eta^{(s \, t)} = \frac{\varepsilon_{m}\lambda \, d_{r}}
     {{\cal A}_{s \, t}}.
     \label{eq:2no19}
\end{eqnarray}
It is worthwhile to note that $\rho^{(s \, t)}$ is positive 
within the frame of Hamiltonian in Eq.~(\ref{eq:2no01}),  
because its numerator is positive, 
as seen from Eqs.~(\ref{eq:2no06}) and (\ref{eq:2no07})~--~(\ref{eq:2no09}).

Two combinations $[n_{2}(x)- s \, n_{1}(x)]$ 
and $[n_{3}(x) - t \, n_{1}(x)]$ in Eq.~(\ref{eq:2no18}) play 
roles of deviation functions for ${\cal N}_{s \, t}(x)$.  
In the paper I, there were some misleading discussions; 
that is, deviation functions presented there are not 
linearly independent as they are, and all of their coefficients 
cannot be settled independently.%%%%
%%%%%%%%%%%%%%%% Footnote begin ********************************
\footnote{In the paper I, a normalization factor 
is denoted by $A_{n \, \ell}$.    
A shortcut to reconstruct linearly independent deviation 
functions is to set $n=\ell$.  Presentations there should be 
corrected as $A_{n \, \ell} \to A_{n \, n}$ and $\rho_{m} \to 0$.  
The old $A_{n \, n}$ corresponds to 
${\cal A}_{2n \, 0}$ in the present paper.}
%%%%%%%%%%%%%%%% Footnote end ********************************

The simple choice of $s$ and $t$ is $(s, \, t)=(0, \, 0)$.  Then, 
the isotropic part is expressed as follows:
\begin{equation}
{\cal N}_{0 \, 0}(x) 
     = n_{1}(x) + n_{2}(x) \, \rho^{(0 \, 0)}
     + n_{3}(x) \, \eta^{(0 \, 0)}.
     \label{eq:2no20}
\end{equation}
This is nothing but the original expression given in 
Eq.~(\ref{eq:2no04}) with $A = {\cal A}_{0 \, 0} = a_{+}$. 
It should be noticed that $A \neq 1$ in principle 
within the frame of our Hamiltonian. 
That is, if the right-handed charged weak current 
or the existence of heavy Majorana neutrino is assumed, 
we have the following expressions 
from Eqs.~(\ref{eq:2no06}) and (\ref{eq:2no07}) respectively:
\begin{eqnarray}
{\cal A}_{0 \, 0} = \left( 1 + \lambda^2 \right) > 1 \hspace{13mm}
     & & \hspace{5mm} \mbox{for the Dirac neutrino case},
     \label{eq:2no20n} \\
  {\cal A}_{0 \, 0} \approx \left( 1 - \overline{u_{e}}^{\, 2} 
     - \overline{u_{\mu}}^{\, 2} \right) \lessapprox 1
     & & \hspace{5mm} \mbox{for the Majorana neutrino case}.
     \label{eq:2no20p}
\end{eqnarray}

The parameters $\rho^{(0 \, 0)}$ and $\eta^{(0 \, 0)}$ are related 
with the weak coupling constants. 
For the Dirac neutrino case, they are
\begin{eqnarray}
\rho^{(0 \, 0)} &\approx&
     \frac{1}{2} \left( \kappa^2 + \eta^{2} \right) > 0,
     \label{eq:2no20q1}\\
  \eta^{(0 \, 0)} &=& 0.
     \label{eq:2no20q2}
\end{eqnarray}
For the Majorana neutrino case, they are
\begin{eqnarray}
\rho^{(0 \, 0)} &\approx&
     \frac{1}{2} \left[ \kappa^2 \,
     \left( \overline{v_{\mu}}^{\, 2} +
          | \, \overline{w_{e \mu}} \, |^{2} \right) +
     \eta^{2} \, 
     \left( \overline{v_{e}}^{\, 2} +
          | \, \overline{w_{e \mu \, h}} \, |^{2} \, \right)
     \right] > 0,
     \label{eq:2no20r1}\\
\eta^{(0 \, 0)} &\approx& \frac{1}{2} \, \lambda \, 
     \mbox{Re} ( \overline{w_{e \mu}}^{\, *} \,\,
     \overline{w_{e \mu \, h}} ).
     \label{eq:2no20r2}
\end{eqnarray}
In these expressions, only the lowest order terms 
are kept by taking ${\cal A}_{0 \, 0} \approx 1$.  
Note that we can get no 
direct information on $\lambda$ from the isotropic spectrum 
${\cal N}_{s \, t}(x)$ in the Dirac neutrino case.  By contrast, 
in the Majorana neutrino case, the parameter $\eta^{(0 \, 0)}$ 
is proportional to $\lambda$.  But the order of magnitude of both 
$\rho^{(0 \, 0)}$ and $\eta^{(0 \, 0)}$ themselves seems to be 
very small, as seen from Eqs.~(\ref{eq:2no20r1}) and (\ref{eq:2no20r2}).

Concerning the relation with the Michel parameter $\rho_{M}$, 
the relevant term in Eq.~(\ref{eq:1no02}) can be reproduced 
from the spectrum ${\cal N}_{s \, t}(x)$ of Eq.~(\ref{eq:2no18}) 
by taking $(s, \, t)=(2, \, 0)$.  That is, we have  
the deviation function $[n_{2}(x) \, - \, 2 \, n_{1}(x)]$ 
and its associated parameter $\rho^{(2 \, 0)}$ as follows;
\begin{equation}
     n_{2}(x)  -  2 \, n_{1}(x)
     \, \simeq \, 2 x^{2} \left( 3 - 4 \, x \right),\hspace{10mm}
     \rho^{(2 \, 0)} = - \frac{2}{3}
     \left( \rho_{M} - \frac{3}{4} \right).
     \label{eq:2no20a}
\end{equation}
It should be noted that the behavior of 
$[n_{2}(x) \, - \, 2 \, n_{1}(x)]$ in 
${\cal N}_{2 \, 0}(x)$ is different from 
$n_{2}(x) \simeq 12 \, x^{2} \left( 1 - x \right)$ 
in ${\cal N}_{0 \, 0}(x)$.  
This will be discussed at the end of \S 4.  

The full expression of the Michel parameterization 
contains another combination $n_{3}(x) \, \eta_{M}$ 
which is omitted in Eq.~(\ref{eq:1no02}) 
because $n_{3}(x)$ is small due to a factor $x_{0}$, 
as seen from Eqs.~(\ref{eq:2no13}) and (\ref{eq:2no16}).~\cite{Fetscher}  
There is a corresponding combination  
$n_{3}(x) \, \eta^{(2 \, 0)}$ in ${\cal N}_{2 \, 0}(x)$.  
These two parameters, $\eta_{M}$ and $\eta^{(2 \, 0)}$, originate 
from different theoretical models for the weak interaction, 
as will be discussed in \S 4. 
But they have the same experimental values phenomenologically. 
We list here the experimental results 
reported by the Particle Data Group:~\cite{Fetscher}
\begin{eqnarray}
\rho^{(2 \, 0)} &=& - (6 \pm 7) \cdot 10^{-4},
     \label{eq:2no20s} \\
  \eta^{(2 \, 0)} &=& \eta_{M} = (1 \pm 24) \cdot 10^{-3}.
     \label{eq:2no20t}
\end{eqnarray}

These parameters are related with the weak coupling constants 
through the same expressions as Eqs.~(\ref{eq:2no20q1}) --
(\ref{eq:2no20r2}) within the lowest order approximation.  
This is because the normalization factor ${\cal A}_{2 \, 0}$ 
takes the following form:
\begin{eqnarray}
 {\cal A}_{2 \, 0} &=& \left( 1 + \lambda^2 +
     \kappa^2 + \eta^{2} \right) > 1 \hspace{7mm} 
     \mbox{for the Dirac neutrino case,}
     \label{eq:2no20u}\\
 {\cal A}_{2 \, 0} &\approx& \left[ 1 - \overline{u_{e}}^{\, 2} -
     \overline{u_{\mu}}^{\, 2} +
     \kappa^2 
     \left( \overline{v_{\mu}}^{\, 2} +
          | \overline{w_{e \mu}} |^{2} \right) +
     \eta^{2}  
     \left( \overline{v_{e}}^{\, 2} +
          | \overline{w_{e \mu \, h}} |^{2} \, \right)
      \right] \approx 1 \nonumber\\
    &&\hspace{45mm}\mbox{for the Majorana neutrino case.}
     \label{eq:2no20v}
\end{eqnarray}

By the way, the spectrum ${\cal N}_{s \, t}(x)$ 
with $s \neq 0$ and/or $t \neq 0$ is derived from 
${\cal N}_{0 \, 0}(x)$ by rearranging the standard function $n_{1}(x)$ 
in it.  Therefore, they should offer the same information on the 
weak coupling constants ($\lambda,\eta$ and $\kappa$).  Indeed, this 
situation is expressed formally as the following identity:
\begin{equation}
{\cal A}_{s \, t} \, {\cal N}_{s \, t}(x) =
     {\cal A}_{0 \, 0} \, {\cal N}_{0 \, 0}(x),
     \label{eq:2no20b}
\end{equation}
as seen from Eqs.~(\ref{eq:2no17}) and (\ref{eq:2no18}).  But 
the normalization factors ${\cal A}_{s \, t}$ and ${\cal A}_{0 \, 0}$ 
themselves do not appear in the data analysis, and furthermore 
the deviation functions are different in the spectrum functions 
${\cal N}_{s \, t}(x)$ and ${\cal N}_{0 \, 0}(x)$.  
We shall examine in \S 3 how to obtain the 
same information mentioned above, when the experimental data 
are analyzed by using different spectrum functions.  For this 
purpose, the notation ${\cal N}_{s \, t}(x)$ will be used to indicate 
the spectrum with $s \neq 0$ and/or $t \neq 0$ hereafter.

There are some useful interrelations between parameters in 
${\cal N}_{0 \, 0}(x)$ and those in ${\cal N}_{s \, t}(x)$. 
The following two are obtained directly from definitions 
in Eqs.~(\ref{eq:2no17}) and (\ref{eq:2no19});
\begin{eqnarray}
\rho^{(0 \, 0)} 
  &=& \frac{1}{\left( 1 - s \, \rho^{(s \, t)} -
     t \, \eta^{(s \, t)} \right)} \rho^{(s \, t)},
     \label{eq:2no22-1} \\
\eta^{(0 \, 0)}
  &=& \frac{1}{\left( 1 - s \, \rho^{(s \, t)} -
     t \, \eta^{(s \, t)} \right)} \eta^{(s \, t)}.
     \label{eq:2no22-2}
\end{eqnarray}
By using these relations, numerical values of 
$\rho^{(0 \, 0)}$ and $\eta^{(0 \, 0)}$ can be estimated 
from the experimental results on 
$\rho^{(2 \, 0)}$ and $\eta^{(2 \, 0)}$. 
There is another identity%%%%
%%%%%%%%%%%%%%%% Footnote begin ********************************
\footnote{We can derive the following relations from 
definitions in Eqs.~(\ref{eq:2no17}) and (\ref{eq:2no19}):
\begin{eqnarray}
{\cal A}_{s \, t} =
     \left( 1 + s \, \rho^{(0 \, 0)} +
     t \, \eta^{(0 \, 0)} \right) \, {\cal A}_{0 \, 0}
  \hspace{3mm} \mbox{or} \hspace{3mm}
     {\cal A}_{0 \, 0} =
     \left( 1 - s \, \rho^{(s \, t)} -
     t \, \eta^{(s \, t)} \right) \, {\cal A}_{s \, t}.
     \nonumber
\end{eqnarray}
\vspace{-4mm}}
%%%%%%%%%%%%%%% Footnote end ************************************
\begin{eqnarray}
     \left( 1 + s \, \rho^{(0 \, 0)} +
     t \, \eta^{(0 \, 0)} \right) \, 
     \left( 1 - s \, \rho^{(s \, t)} -
     t \, \eta^{(s \, t)} \right) = 1,
     \label{eq:2no23-1}
\end{eqnarray}
from which we can derive the inverse relations 
to express $\rho^{(s \, t)}$ and $\eta^{(s \, t)}$ 
in terms of $\rho^{(0 \, 0)}$ and $\eta^{(0 \, 0)}$. 

It can be shown by using these identities 
that the relation in Eq.~(\ref{eq:2no20b}) 
can be expressed as follows:
\begin{eqnarray}
{\cal N}_{s \, t}(x) = 
     \frac{1}{\left( 1 + s \, \rho^{(0 \, 0)} +
     t \, \eta^{(0 \, 0)} \right)}
      \, {\cal N}_{0 \, 0}(x).
     \label{eq:2no23-2}
\end{eqnarray}
This implies that the spectrum function ${\cal N}_{s \, t}(x)$ 
as well as parameters ($\rho^{(s \, t)}, \, \eta^{(s \, t)}$) 
can be obtained from knowledge about 
${\cal N}_{0 \, 0}(x)$ and ($\rho^{(0 \, 0)}, \eta^{(0 \, 0)}$), 
and vice versa.

\subsection{Anisotropic part of the spectrum: ${\cal P}(x)$}

Next, let us consider the anisotropic part ${\cal P}(x)$ 
in Eq.~(\ref{eq:2no05}).  We take the standard 
function $p_{1}(x)$ as a base of rearrangement.  In order to see 
the relation with Michel parameterization, we define a common factor 
${\cal B}_{u}$ by using the coefficient of $p_{2}(x)$;
\begin{eqnarray}
{\cal B}_{u} = a_{-} + u \, ( k_{- \, c} +
     \varepsilon_{m} k_{- \, m} ),
     \label{eq:2no21}
\end{eqnarray}
where $u$ is some arbitrary number.%%%%
%%%%%%%%%%%%%%%% Footnote begin ********************************
\footnote{In the paper I, a common factor is 
denoted by $B_{n \, \ell}$.  
Presentations there should be corrected 
as $B_{n \, \ell} \to B_{n \, n}$ and $\delta_{m} \to 0$.  
Note that $B_{n \, n}$ corresponds to ${\cal B}_{2n}$ in the present paper.}
%%%%%%%%%%%%%%%% Footnote end ********************************
Then, the anisotropic spectrum is denoted as ${\cal P}_{s \, t \, u}(x)$:
\begin{equation}
{\cal P}_{s \, t \, u}(x) = \xi^{(s \, t \, u)} \, \left\{ p_{1}(x) +
     \bigl[ p_{2}(x) - u \, p_{1}(x) \bigr]\, \delta^{(u)}
     \right\},
     \label{eq:2no22}
\end{equation}
where parameters are defined as
\begin{eqnarray}
\xi^{(s \, t \, u)} = \frac{{\cal B}_{u}}{{\cal A}_{s \, t }}, 
     \hspace{20mm}
     \delta^{(u)} =
     \frac{k_{- \, c} + \varepsilon_{m} \, k_{- \, m}}{{\cal B}_{u}}.
\label{eq:2no23}
\end{eqnarray}

For the simple choice	$(s,\,t,\,u)=(0,\,0,\,0)$, 
we have ${\cal A}_{0 \, 0} = a_{+}$ and ${\cal B}_{0} = a_{-}$.  
Then, the anisotropic spectrum is expressed as
\begin{equation}
{\cal P}_{0 \, 0 \, 0}(x) = \xi^{(0 \, 0 \, 0)} \,
     \left[ p_{1}(x) + p_{2}(x) \, \delta^{(0)} \right] .
     \label{eq:2no26}
\end{equation}
This ${\cal P}_{0 \, 0 \, 0}(x)$ is identical with 
Eq.~(\ref{eq:2no05}) by taking 
$\xi^{(0 \, 0 \, 0)} = (a_{-}/a_{+})$.

The parameters $\xi^{(0 \, 0 \, 0)}$ and $\delta^{(0)}$ 
are related with the weak coupling constants. 
For the Dirac neutrino case, they are
\begin{equation}
\xi^{(0 \, 0 \, 0)} = \frac{\left( 1 - \lambda^2 \right)}
     {\left( 1 + \lambda^2 \right)},
     \hspace{10mm}
\delta^{(0)} \approx
     \frac{1}{2} \left( \kappa^2 - \eta^{2} \right). 
     \label{eq:2no45}
\end{equation}
For the Majorana neutrino case, they are
\begin{equation}
\xi^{(0 \, 0 \, 0)} \approx 1-
     2\lambda^2\overline{v_{e}}^{\, 2}\overline{v_{\mu}}^{\, 2}
     \approx 1, \hspace{5mm}
\delta^{(0)} \approx
     \frac{1}{2} \left[ \kappa^2 
     \left( \overline{v_{\mu}}^{\, 2} +
          | \overline{w_{e \mu}} |^{2} \right) -
     \eta^{2} 
     \left( \overline{v_{e}}^{\, 2} +
          | \overline{w_{e \mu \, h}} |^{2} \right) \right].
     \label{eq:2no46}
\end{equation}
Here, only the leading terms are given 
for the deviation from the standard model.

On the other hand, the Michel parameterization 
in Eq.~(\ref{eq:1no03}) can be reproduced from 
${\cal P}_{s \, t \, u}(x)$ by choosing $(s,\,t,\,u)=(2,\,0,\,6)$; 
that is, the following correspondences are obtained:
\begin{equation}
     p_2(x)-6p_1(x) \, \simeq \, 6x^2(3-4x),\hspace{5mm}
     \xi^{(2 \, 0 \, 6)} = \xi_{M},\hspace{5mm} 
     \delta^{(6)} =\frac{2}{9}\left(\frac{3}{4}- \delta_{M}\right).
     \label{eq:2no47}
\end{equation}
The experimental results reported by the Particle Data 
Group~\cite{Fetscher} are as follows:
\begin{eqnarray}
\left| \xi^{(2 \, 0 \, 6)} \, P_{\mu} \right|
     &=& 1.0027 \pm 0.0079 \pm 0.0030,
     \label{eq:2no48} \\
  \delta^{(6)} &=& (1.1 \pm 2.7) \cdot 10^{-4}.
     \label{eq:2no49}
\end{eqnarray}
Here $P_{\mu}$ stands for the longitudinal polarization of the 
muon introduced in Eq.~(\ref{eq:1no01}).  The common 
parameter $\xi^{(2 \, 0 \, 6)}$ is related with 
the weak coupling constants:
\begin{eqnarray}
\xi^{(2 \, 0 \, 6)} &=&
     \frac{1 - \lambda^2 + 3 \, (\kappa^2 - \eta^{2})}
     {1 + \lambda^2 + \kappa^2 + \eta^{2}}
     \hspace{5mm} \mbox{for the Dirac neutrino case},
     \label{eq:2no50} \\
   \xi^{(2 \, 0 \, 6)} &\approx&
     1+2\kappa^2(\overline{v_{\mu}}^2+ \overline{w_{e \mu}}^2)
     -4\eta^2(\overline{v_{e}}^2+ \overline{w_{e \mu h}}^2)
     \approx 1
    \nonumber\\
     &&\hspace{39mm} \mbox{for the Majorana neutrino case}.
     \label{eq:2no51}
\end{eqnarray}
The parameter $\delta^{(6)}$ is expressed in the same form 
as $\delta^{(0)}$ in Eqs.~(\ref{eq:2no45}) or (\ref{eq:2no46}) 
if only the leading terms are kept 
for the deviation from the standard model.  

Parameters in ${\cal P}_{s \, t \, u}(x)$ and 
${\cal P}_{0 \, 0 \, 0}(x)$ satisfy the following identities:
\begin{eqnarray}
\xi^{(0 \, 0 \, 0)} &=& 
     \frac{\left[ 1 - u  \delta^{(u)} \right]}
     {\left[ 1 - s \, \rho^{(s \, t)} -
     t \, \eta^{(s \, t)} \right]}\xi^{(s \, t \, u)}, 
     \label{eq:2no52} \\
 \delta^{(0)} &=& \frac{1}
     {\left[ 1 - u \, \delta^{(u)} \right]} \, \delta^{(u)}.
     \label{eq:2no53}
\end{eqnarray}
The inverse relations are obtained by using the following identity.
\begin{eqnarray}
\left(1 + u \, \delta^{(0)} \right)
     \left( 1 - u \, \delta^{(u)} \right) = 1.
     \label{eq:2no54}
\end{eqnarray}
All these relations are derived from the definitions in 
Eqs.~(\ref{eq:2no21}) and (\ref{eq:2no23}).  Also we can 
confirm the following relation,
\begin{eqnarray}
{\cal P}_{s \, t \, u}(x) =
     \frac{1}{\left( 1 + s \, \rho^{(0 \, 0)} +
     t \, \eta^{(0 \, 0)} \right)}
      \, {\cal P}_{0 \, 0 \, 0}(x).
     \label{eq:2no55}
\end{eqnarray}
Note that this relation is independent of $u$ 
introduced in Eq.~(\ref{eq:2no21}).

Finally, it is useful to note that we have $\kappa = \eta$ 
if the $SU(2)_{L} \times SU(2)_{R} \times U(1)$ model is assumed, 
and that the $\delta^{(u)}$ parameter becomes simpler:
\begin{eqnarray}
\delta^{(u)} &=& 0 \hspace{35mm} \mbox{for the Dirac neutrino case},
     \label{eq:2no56} \\
\delta^{(u)} &\simeq& \frac{\eta^{2}}{2}
     \left[ \left( \overline{v_{\mu}}^{\, 2} -
     \overline{v_{e}}^{\, 2} \right) +
     \left( | \, \overline{w_{e \mu}} \, |^{2} -
     | \, \overline{w_{e \mu \, h}} \, |^{2} \right)
     \right] \ll 1
     \nonumber \\
  & & {} \hspace{37mm} \mbox{for the Majorana neutrino case}.
     \label{eq:2no57}
\end{eqnarray}

%%%%%%%%%%%%%%%%%%%%%%%%%%%%%%%%%%%%%%%%%%%%%%%%%%%%%%%%%%%%%%%
%%%%%%%%%%%%%%%%%          Section 3         %%%%%%%%%%%%%%%%%%
%%%%%%%%%%%%%%%%%%%%%%%%%%%%%%%%%%%%%%%%%%%%%%%%%%%%%%%%%%%%%%%
\section{\label{sec:three}Method of least squares}

Let us find the condition to get the same results for the weak 
coupling constants ($\lambda, \, \eta$ and $\kappa$), when different 
spectrum functions are adopted in the data analysis.  Also, the 
method is examined to include the QED radiative corrections.  We 
shall first concentrate our discussion on the isotropic part 
${\cal N}_{s \, t}(x)$.  It is easy to extend our 
treatment to the full spectrum where the anisotropic part 
${\cal P}_{s \, t \, u}(x)$ is taken into consideration.  

\subsection{Analysis of the isotropic part of the spectrum:\ ${\cal N}(x)$}

We shall use the method of least squares in the data analysis.  The 
QED radiative correction is not included for a while.  In the 
case where ${\cal N}_{s \, t}(x)$ is used, the unknown 
parameters ($\rho^{(s \, t)}$ and $\eta^{(s \, t)}$) are settled 
as some values, at which the following $\chi_{\, s \, t}^{2}$ 
takes a minimum:
\begin{equation}
     \chi_{\, s \, t}^{2} = \sum_{i}\frac{1}{\sigma_{i}^{\; 2}}
     \Bigl[ {\cal E}(x_{i}) - c_{s \, t} \, 
     {\cal N}_{s \, t}(x_{i}) \Bigr]^2.
     \label{eq:3no1}
\end{equation}
The summation over $i$ runs over all measuring points $x_{i}$.  
The notation ${\cal E}(x_{i})$ stands for an 
experimental datum at $x_{i}$, 
and $\sigma_{i}$ is its experimental error. 
The global normalization constant $c_{s \, t}$ is introduced 
to adjust the theoretical values to the experimental data, 
so that the minimum point of $\chi_{\, s \, t}^2$ is sought 
under the variation of $c_{s \, t}$ as well as the parameters.

By requiring that $\chi_{\, s \, t}^{2}$ takes a minimum, 
a set of analytical solutions is obtained for 
parameters ($c_{s \, t} $, $c_{s \, t} \, \rho^{(s \, t)}$ 
and $c_{s \, t} \, \eta^{(s \, t)}$).  They are known as the 
Cramers' formula for the system of linear 
equations,~\cite{ParticleData} because these parameters appear 
linearly in $c_{s \, t} \, {\cal N}_{s \, t}(x)$.  
If the spectrum function ${\cal N}_{0 \, 0}(x)$ is adopted 
and $\chi_{\, 0 \, 0}^{2}$ is required to have a minimum, 
parameters ($c_{0 \, 0}$, $c_{0 \, 0} \, \rho^{(0 \, 0)}$ 
and $c_{0 \, 0} \, \eta^{(0 \, 0)}$) are expressed by 
another set of analytical solutions.  
These two sets of solutions indicate that 
two global normalization constants 
$c_{s \, t}$ and $c_{0 \, 0}$ satisfy the relation:
\begin{equation}
      \left(\frac{c_{s \, t}}{c_{0 \, 0}}\right) 
     = \left( 1 + s \, \rho^{(0 \, 0)} +
     t \, \eta^{(0 \, 0)} \right).
     \label{eq:3no2}
\end{equation}
It can be confirmed further that, under this relation, 
solutions for $(\rho^{(s \, t)}, \, \rho^{(0 \, 0)})$ and 
$(\eta^{(s \, t)}, \, \eta^{(0 \, 0)})$ are consistent with 
their interrelations in Eqs.~(\ref{eq:2no22-1}) 
and (\ref{eq:2no22-2}), respectively.  
Therefore, the relation in Eq.~(\ref{eq:3no2}) 
will be called the equivalent condition hereafter.  

If we combine Eq.~(\ref{eq:3no2}) with  Eq.~(\ref{eq:2no23-2}), 
the following equality is obtained:
\begin{equation}
c_{s \, t} \, {\cal N}_{s \, t}(x) =
     c_{0 \, 0} \, {\cal N}_{0 \, 0}(x).
     \label{eq:3no3}
\end{equation}
This means that the $\chi^{2}$-values are the same 
for different spectrum functions ${\cal N}_{s \, t}(x)$ and 
${\cal N}_{0 \, 0}(x)$; that is,
\begin{equation}
    \chi_{\, s \, t}^{2} = \chi_{\,0 \, 0}^{2}.
     \label{eq:3no4}
\end{equation}

In summary, the parameters ($\rho^{(0 \, 0)}$ and $\eta^{(0 \, 0)}$) 
are settled experimentally, when the $\chi_{\, 0 \, 0}^2$-value becomes 
a minimum.  The consistency between independent data analyses with use 
of ${\cal N}_{0 \, 0}(x)$ and ${\cal N}_{s \, t}(x)$ is guaranteed by 
the equivalent condition for the global normalization constants 
($c_{0 \, 0}$ and $c_{s \, t}$).  Then, $\chi_{\, s \, t}^2$ becomes 
to be equal to $\chi_{\, 0 \, 0}^2$, when their parameters satisfy 
interrelations in Eqs.~(\ref{eq:2no22-1}) and (\ref{eq:2no22-2}).  
But, we should be aware that the equivalent condition 
is on the delicate balance of the global 
normalization constants, $c_{0 \, 0}$ and $c_{s \, t}$, 
because their difference is very slight due to smallness of 
$\rho^{(0 \, 0)}$ and $\eta^{(0 \, 0)}$.

Now let us examine how to take account of the QED radiative 
correction in the analysis of the spectrum.  As the first step, 
we shall consider the case where the data analysis is 
performed by assuming the standard model.  
Then, we try to find a minimum value of 
the following $X_{\, \mathrm{sm}}^2$:
\begin{equation}
X_{\, \mathrm{sm}}^{2}= \sum_{i}\frac{1}{\sigma_{i}^{\; 2}}
     \Bigl| {\cal E}(x_{i}) - c_{\mathrm{sm}} \, 
     \left[ n_{1}(x_{i}) +  f(x_{i}) \right] \Bigr|^2,
     \label{eq:3no5}
\end{equation}
where $f(x)$ stands for the QED radiative correction 
associated with the standard function $n_{1}(x)$ in 
Eq.~(\ref{eq:2no11}).~\cite{Arbuzov}%%%
%%%%%%%%%%%%%%%% Footnote begin ********************************
\footnote{Relation between our $f(x)$ and 
$F(x)$ by Arbuzov~\cite{Arbuzov} is 
$f(x) = \Big[1 + (m_{e}/m_{\mu})^{2} \Big]^{-4} F(x)
  - {\cal N}_{2 \, 0}(x).$}
%%%%%%%%%%%%%%%% Footnote end ********************************
Note that the unknown parameter in $X_{\, \mathrm{sm}}^{2}$ 
is only the global normalization constant $c_{\mathrm{sm}}$.

If the effect due to the weak coupling constants ($\lambda, \, 
\eta$ and $\kappa$) is considered, then the above $X_{\, \mathrm{sm}}^2$ 
is modified.  The standard function $n_{1}(x)$ is replaced 
by the spectrum function ${\cal N}_{0 \, 0}(x)$ or 
${\cal N}_{s \, t}(x)$ in Eq.~(\ref{eq:2no18}), 
but the QED radiative correction $f(x)$ is received no influence 
because of the consistency of the approximation.  
Thus, it is appropriate to introduce 
the following form instead of $X_{\, \mathrm{sm}}^2$:
\begin{equation}
X_{\, s \, t}^{2} = \sum_{i}\frac{1}{\sigma_{i}^{\; 2}}
     \left[ {\cal E}(x_{i}) - c_{s \, t} \, 
     {\cal N}_{s \, t}(x_{i}) - c_{\mbox{\tiny $R$}} \, f(x_{i}) \right]^2.
     \label{eq:3no6}
\end{equation}
Here it is understood that the notation $(s, \, t)$ includes 
the case of $(s = 0, \, t = 0)$.  Two new parameters $c_{s \, t}$ and 
$c_{\mbox{\tiny $R$}}$ are different from $c_{\mathrm{sm}}$ in 
$X_{\, \mathrm{sm}}^{2}$, because ${\cal N}_{s \, t}(x)$ is present 
in place of $n_{1}(x)$.  Note that this $c_{\mbox{\tiny $R$}}$ is 
independent of $s$ and $t$.  This can be confirmed by comparing two 
sets of analytical solutions for parameters ($c_{0 \, 0}, \, 
\rho^{(0 \, 0)}, \, \eta^{(0 \, 0)}, \, c_{\mbox{\tiny $R$}}$) and 
($c_{s \, t}, \, \rho^{(s \, t)}, \, \eta^{(s \, t)}, \, 
c_{\mbox{\tiny $R$}}$).  Each of these sets is obtained by requiring 
that $X_{\, 0 \, 0}^{2}$ or $X_{\, s \, t}^{2}$ has its minimum.

The equivalent condition in Eq.~(\ref{eq:3no2}) is derived again 
for these new global normalization constants, $c_{0 \, 0}$ and 
$c_{s \, t}$, introduced in Eq.~(\ref{eq:3no6}).  Then, it can be 
proved under this equivalent condition that the 
$\chi^{2}$-values are the same for different spectrum functions 
${\cal N}_{0 \, 0}(x)$ and ${\cal N}_{s \, t}(x)$ 
whose parameters satisfy interrelations in Eq.~(\ref{eq:2no22-1}) 
for $(\rho^{(0 \, 0)}, \, \rho^{(0 \, 0)})$ 
and those in Eq.~(\ref{eq:2no22-2}) for 
$(\eta^{(0 \, 0)}, \, \eta^{(s \, t)})$:
\begin{equation}
    X_{\, s \, t}^{2} = X_{\, 0 \, 0}^{2}.
     \label{eq:3no7}
\end{equation}

It is worthwhile to make the following three comments.  
First we shall examine whether the special choice of the spectrum 
function is preferable in the actual numerical analysis.  
For this purpose, it is useful to estimate the $x$-dependence 
of the function $\Delta(x)$ defined by
\begin{equation}
\Delta(x) = {\cal E}(x) - c_{\mathrm{sm}} \, 
     \left[ n_{1}(x) +  f(x) \right],
     \label{eq:3no5-1}
\end{equation}
where $c_{\mathrm{sm}}$ is fixed such that $X_{\, \mathrm{sm}}^{2}$ 
in Eq.~(\ref{eq:3no5}) takes a minimum.  If this $\Delta(x)$ 
shows any $x$-dependence clearly, then we may choose such a value 
of $s$ that $\Delta(x)$ is roughly proportional to the deviation function 
$[n_2(x) - s \, n_1(x)]$ in Eq.~(\ref{eq:2no18}).  
However, it is imaginable 
that $\Delta(x)$ does not show any clear $x$-dependence, because of 
experimental errors.  If this is the case, then it may be preferable 
to adopt ${\cal N}_{0 \, 0}(x)$ in Eq.~(\ref{eq:2no20}), 
because parameters are related with weak coupling constants 
in simpler forms.  As a conclusion, we propose to use $X_{\, 0 \, 0}^{2}$ 
in the actual data analysis for its simplicity.

The next comment is that the precise determination of 
the $X_{\, \mathrm{sm}}^{2}$-value itself defined 
in Eq.~(\ref{eq:3no5}) is interesting.  
This is because the large deviation from the standard model 
cannot be expected, especially for the Majorana neutrino case. 
This will be discussed at the last paragraph in \S 4.

The final comment is that, in contrast to $X_{\, s \, t}^{2}$ 
in Eq.~(\ref{eq:3no6}), the following definition
\begin{equation}
Y_{\, s \, t}^{2} = \sum_{i}\frac{1}{\sigma_{i}^{\; 2}}
     \Bigl| {\cal E}(x_{i}) - c_{s \, t} \,
     [{\cal N}_{s \, t}(x_{i}) + f(x_{i})] \Bigr|^2,
     \label{eq:3no8}
\end{equation}
is not appropriate theoretically, because it leads to 
the inequality $Y_{\, s \, t}^{2} \ne Y_{\, 0 \, 0}^{2}$.

%%%%%%%%%%%%%%%%%%%%%%%%%%%%%%%%%%%%%%%%%%%%%%%%%%%%%%%%%%%%%%%
\subsection{Analysis of the full spectrum:\ ${\cal D}(x)$}
%%%%%%%%%%%%%%%%%%%%%%%%%%%%%%%%%%%%%%%%%%%%%%%%%%%%%%%%%%%%%%%

In the extended form of the parameterization, 
the full spectrum in Eq.~(\ref{eq:2no02}) is expressed as follows:
\begin{equation}
{\cal D}_{s \, t \, u}(x, \, \theta) = \Bigl[ {\cal N}_{s \, t}(x) +
      P_{\mu} \, \cos \theta \, {\cal P}_{s \, t \, u}(x) \, \Bigr],
      \label{eq:3no10}
\end{equation}
where ${\cal N}_{s \, t}(x)$ and ${\cal P}_{s \, t \, u}(x)$ are 
given by Eqs.~(\ref{eq:2no18}) and (\ref{eq:2no22}), respectively. 
The method of least squares can be applied similarly 
to the isotropic part ${\cal N}_{s \, t}(x)$.

We summarize the essential points for the case with no radiative 
correction.  The new $\chi_{\, s \, t \, u}^{2}$ is defined as follows:
\begin{equation}
     \chi_{\, s \, t \, u}^{2} = 
     \sum_{i, \, j}\frac{1}{\sigma_{i \, j}^{\; 2}}
     \Bigl[ {\cal E}(x_{i}, \, \theta_{j}) - d_{s \, t} \, 
     {\cal D}_{s \, t \, u}(x_{i}, \, \theta_{j}) \Bigr]^2,
     \label{eq:3no11}
\end{equation}
where $x_{i}$ and $\theta_{j}$ are a set of observed quantities 
at one measuring point.  We have the analytical solutions for 
two new parameters 
($\xi^{(s \, t \, u)}$ and $\delta^{(u)}$) in addition to 
three old ones ($d_{s \, t}, \, \rho^{(s \, t)}$ and 
$\eta^{(s \, t)}$) by requiring that $\chi_{\, s \, t \, u}^{2}$ 
takes a minimum.  The corresponding solutions are 
also obtained by treating $\chi_{\, 0 \, 0 \, 0}^{2}$.  
It should be noted that the global normalization constant 
$d_{s \, t}$ here depends only on $s$ and $t$, 
because it is settled as a coefficient for ${\cal N}_{s \, t}(x)$ 
of ${\cal D}_{s \, t \, u}(x, \, \theta)$ in the method of least squares.  
Then, it can be verified not only that the global normalization constants 
satisfy the similar equivalent condition to the one in 
Eq.~(\ref{eq:3no2}), but also that other four parameters are 
consistent with 
interrelations in Eqs.~(\ref{eq:2no22-1}), (\ref{eq:2no22-2}), 
(\ref{eq:2no52}) and (\ref{eq:2no53}).  
After all, due to this equivalent condition together with relations in 
Eqs.~(\ref{eq:2no23-2}) and (\ref{eq:2no55}), we have the identity 
$d_{s \, t}{\cal D}_{s \, t \, u}(x, \, \theta)
=d_{0 \, 0}{\cal D}_{0 \, 0 \, 0}(x, \, \theta)$ 
and subsequently the equality 
\begin{equation}
     \chi_{\, s \, t \, u}^{2} = \chi_{\, 0 \, 0 \, 0}^{2}.
     \label{eq:3no12}
\end{equation}

In the case where the radiative QED effect is taken into consideration, 
we modify $\chi_{\, s \, t \, u}^{2}$ and define the following 
$Z_{\, s \, t \, u}^{2}$ which satisfies both the 
consistency conditions for parameters and the equality 
$Z_{\, s \, t \, u}^{2} = Z_{\, 0 \, 0 \, 0}^{2}$:
\begin{equation}
     Z_{\, s \, t \, u}^{2} = 
     \sum_{i, \, j}\frac{1}{\sigma_{i \, j}^{\,\, 2}}
     \Bigl[ {\cal E}(x_{i}, \, \theta_{j}) - d_{s \, t} \, 
     {\cal D}_{s \, t \, u}(x_{i}, \, \theta_{j}) -
     d_{\mbox{\tiny $R$}} \, F(x_{i}, \, \theta_{j}) \Bigr]^2,
     \label{eq:3no13}
\end{equation}
where a new parameter $d_{\mbox{\tiny $R$}}$ corresponds 
to $c_{\mbox{\tiny $R$}}$ in Eq.~(\ref{eq:3no6}) and
\begin{equation}
F(x, \, \theta) = \bigl[ f(x) + P_{\mu} \, \cos \theta \, g(x)
      \, \bigr].
      \label{eq:3no14}
\end{equation}
Here $g(x)$ stands for the QED radiative correction 
associated with the anisotropic standard function 
$p_{1}(x)$ in Eq.~(\ref{eq:2no14}).~\cite{Arbuzov}%%%
%%%%%%%%%%%%%%%% Footnote begin ********************************
\footnote{Relation between our $g(x)$ and $G(x)$ 
by Arbuzov~\cite{Arbuzov} is 
$g(x) = - \Big[ 1 + (m_{e}/m_{\mu})^{2} \Big]^{-4} G(x)
- {\cal P}_{2 \, 0}(x)$.}
%%%%%%%%%%%%%%%% Footnote end ********************************

Corresponding to $X_{\, \mathrm{sm}}^{2}$ for the isotropic part 
of the standard model, the following $Z_{\, \mathrm{sm}}^{2}$ with a 
global normalization constant $d_{\mathrm{sm}}$ is defined for 
the full spectrum:
\begin{equation}
     Z_{\, \mathrm{sm}}^{2} =
     \sum_{i, \, j}\frac{1}{\sigma_{i \, j}^{\,\, 2}}
     \Bigl| {\cal E}(x_{i}, \, \theta_{j}) -
     d_{\mathrm{sm}} \,
     \left[ D_{\mathrm{sm}}(x_{i}, \, \theta_{j}) +
     F(x_{i}, \, \theta_{j}) \right] \Bigr|^2,
     \label{eq:3no15}
\end{equation}
where
\begin{equation}
D_{\mathrm{sm}}(x, \, \theta) =
     \bigl[ n_{1}(x) + P_{\mu} \, \cos \theta \, p_{1}(x) \, \bigr].
     \label{eq:3no16}
\end{equation}
It is interesting to compare minima of $Z_{\, \mathrm{sm}}^{2}$ 
and $Z_{\, 0 \, 0 \, 0}^{2}$ in order to know directly the extent 
of departure from the standard model.

%%%%%%%%%%%%%%%%%%%%%%%%%%%%%%%%%%%%%%%%%%%%%%%%%%%%%%%%%%%%%%%
%%%%%%%%%%%%%%%%%          Section 4         %%%%%%%%%%%%%%%%%%
%%%%%%%%%%%%%%%%%%%%%%%%%%%%%%%%%%%%%%%%%%%%%%%%%%%%%%%%%%%%%%%
\section{\label{sec:four}Discussion}

Let us consider the possible method to determine 
whether the neutrino is of the Dirac or Majorana type.  
It is offered by observing the $\eta^{(s \, t)}$ parameter 
which is zero or nonzero depending on the Dirac 
or Majorana neutrino within the frame of the gauge theory, 
as seen from Eqs.~(\ref{eq:2no20q2}) and (\ref{eq:2no20r2}).%%%%%
%%%%%%%%%%%%%% Footnote begin %%%%%%%%%%%%%%%%%%%%%%%%%%%%%%%%%%%
\footnote{This difference is independent of the choice of the 
normalization factor $A$.  It should be noted on this point that there 
are some misleading discussions in the paper I.  It is not 
correct to say that there is some difference between the Dirac 
and Majorana cases by choosing $A$.}
%%%%%%%%%%%%%% Footnote end %%%%%%%%%%%%%%%%%%%%%%%%%%%%%%%%%%%%%
  Meanwhile, the Michel parameter $\eta_{M}$ has been popular as 
a measure to show the deviation from the standard model.  It 
corresponds to $\eta^{(2 \, 0)}$, as mentioned in  
Eq.~(\ref{eq:2no20t}).  But, the $\eta^{(s \, t)}$ term is 
defined for the Majorana neutrino case within the frame of 
gauge theory, while the $\eta_{M}$ term comes from the 
interference between the $(V \pm A)$ and $(S \pm P)$ (or $T$) 
forms even for the massless Dirac neutrino case.~\cite{Fetscher}  
Anyway, the observation of this $\eta$ parameter 
indicates the deviation from the standard model.

It is well known that the detection of this $\eta$ parameter 
is very difficult experimentally.  One of its reasons is that 
the contribution of the relevant deviation function $n_{3}(x)$ 
is small, because it is proportional to the small value of $x_{0}$, 
as shown in Eq.~(\ref{eq:2no16}).  We may avoid this weak point by 
considering the $\tau$-decay:
\begin{equation}
	\tau^+ \to \mu^{+} + \nu_{\mu} + \overline{\nu_{\tau}}.
     \label{eq:4no1}
\end{equation}
All formula in the previous sections can be 
applied to the $\tau$-decay by the replacement of both 
($m_{\mu} \to m_{\tau}$) and ($m_{e} \to m_{\mu}$).  
Then, the value of $x_{0} = m_{e}/W \simeq 0.01$ 
is shifted up to $x_0(\tau)=m_{\mu}/W_{\tau} \simeq 0.12$ 
where $W_{\tau}=(m_{\tau}^2+m_{\mu}^2)/2m_{\tau}$.  Meanwhile, 
the second reason for the difficulty to detect 
the $\eta^{(s \, t)}$ parameter is its smallness 
within the gauge theory, as seen from Eqs.~(\ref{eq:2no20q2}) and 
(\ref{eq:2no20r2}).  Its rough estimate was discussed in 
\S 4.2 of the paper I.  As a conclusion, the muon decay cannot be 
used to discriminate between the Majorana 
and Dirac neutrino cases in reality.

Next, let us examine the order of magnitude of the 
normalization factor $A$.  In the traditional model which 
has been used to analyze the experimental results, it has been 
assumed that $A = {\cal A}_{2 \, 0} = 1$.~\cite{Kuno2001}  However, 
in our model for the Dirac neutrino, as we know  
${\cal A}_{0 \, 0} = \left( 1 + \lambda^2 \right) > 1$ 
from Eq.~(\ref{eq:2no20n}), we need some information on 
$\lambda^2$.  We may say that $\lambda^2 < \mathrm{O}(10^{-3})$ 
from the reported data in 
Eqs.~(\ref{eq:2no48})~--~(\ref{eq:2no50}) and 
(\ref{eq:2no20s}).  In the Majorana neutrino case, there 
is no definite information at present, although it is imagined from 
Eq.~(\ref{eq:2no20p}) that the deviation from 
${\cal A}_{0 \, 0} = 1$ is very small.

Finally, we would like to comment on the data of the Michel 
parameter $\rho_{M}$.  The mean value of $\rho^{(2 \, 0)}$ 
obtained from $\rho_{M}$ is $\rho^{(2 \, 0)} = - 6 \cdot 10^{-4}$, 
as shown in Eq.~(\ref{eq:2no20s}).  This mean value is negative, 
although it can be positive within experimental uncertainty.  
From the theoretical side, it is predicted to be positive 
within the frame of Hamiltonian in Eq.~(\ref{eq:2no01}), 
as seen from Eqs.~(\ref{eq:2no20q1}) or (\ref{eq:2no20r1}) 
and (\ref{eq:2no22-1}).  There is a possibility that this difference 
comes from some ambiguity in the data analysis.  This is because 
the consistency of the data analysis depends delicately on the 
equivalent condition in Eq.~(\ref{eq:3no2}) and the means to treat 
the QED radiative correction, as mentioned in \S 3.  
Under these circumstances, it is of interest to compare 
experimental results for $\rho^{(2 \, 0)}$ and $\rho^{(0 \, 0)}$, 
which are determined  by using ${\cal N}_{2 \, 0}(x)$ 
and ${\cal N}_{0 \, 0}(x)$, respectively.  
These parameters should satisfy the interrelation in 
Eq.~(\ref{eq:2no22-1}) and have the same signature.  
In this connection, the evaluation of 
$X_{\, 0 \, 0}^{2}$ and $X_{\, 2 \, 0}^{2}$ is of importance, 
because they are equal theoretically.  
It is also interesting to compare them with $X_{\, \mathrm{sm}}^{2}$ 
in Eq.~(\ref{eq:3no5}), 
because the large deviation from the standard model cannot be expected, 
especially for the Majorana neutrino case.

%%%%%%%%%%%%%%%%%%%%%%%%%%%%%%%%%%%%%%%%%%%%%%%%%%%%%%%%%%%%%%%
%%%%%%%%%%%%%%%%%          Appendix A        %%%%%%%%%%%%%%%%%%
%%%%%%%%%%%%%%%%%%%%%%%%%%%%%%%%%%%%%%%%%%%%%%%%%%%%%%%%%%%%%%%
\appendix
\section{Polarization of positron}\label{sc:AT1}

Since parameters are defined in somewhat different forms from the 
paper I, we shall list expressions for the 
polarization of an emitted positron.

The differential decay rate is expressed as follows;
\begin{equation}
\frac{d^2 \Gamma}{d x \, d \cos \theta}
     = \frac{1}{2} \Gamma_{W} \, A \, {\cal D}(x, \, \theta) \,
     \left[ 1 + \vec{P_{e}}(x, \, \theta) \cdot 
     \hat{\zeta} \right],
     \label{eq:Ano01}
\end{equation}
where the vector $\vec{P_{e}}(x, \, \theta)$ is a polarization vector 
of $e^{+}$, and $\hat{\zeta}$ is the directional vector of the 
measurement of the $e^{+}$ spin polarization. The decay plane is
defined by the momentum direction ($\vec{p_{e}}$) of $e^{+}$ and 
the muon polarization vector ($\vec{P_{\mu}}$).

Three components of the $e^{+}$ spin polarization vector are defined 
as~\cite{Fetscher}
\begin{equation}
  \vec{P}_{e}(x, \, \theta) 
     = P_{L}(x, \, \theta) \hat{p_{e}}
     + P_{T1}(x, \, \theta)
       \frac{(\hat{p_{e}} \times \vec{P_{\mu}}) \times \hat{p_{e}}}
       {|(\hat{p_{e}} \times \vec{P_{\mu}}) \times \hat{p_{e}}|}
     + P_{T2}(x, \, \theta)
       \frac{\hat{p_{e}} \times \vec{P_{\mu}}}
       {|\hat{p_{e}} \times \vec{P_{\mu}}|}.
  \label{eq:Ano02}
\end{equation}
The explicit expressions of these $P_{L}(x, \, \theta)$, 
$P_{T1}(x, \, \theta)$, and $P_{T2}(x, \, \theta)$ are presented in 
terms of parameters defined in \S 2 of the present paper.  For 
simplicity, they are listed only for 
the simple form with both the normalization 
factor ${\cal A}_{0 \, 0}$ and the common factor ${\cal B}_{0}$.  
Also, the radiative corrections are not included here.\cite{Arbuzov}

%%%%%%%%%%%%%%%%%%%% A-1 %%%%%%%%%%%%%%%%%%%%%%%%%%%%%%%%%%%%%%%
\subsection{Longitudinal polarization: $P_{L}(x, \, \theta)$}
%%%%%%%%%%%%%%%%%%%% A-1 %%%%%%%%%%%%%%%%%%%%%%%%%%%%%%%%%%%%%%%

It is convenient to separate the isotropic and anisotropic 
distributions of $e^{+}$ with respect to the muon polarization 
vector $\vec{P}_{\mu}$, namely,
\begin{eqnarray}
P_{L}(x, \, \theta)
  = \frac{Q(x) + P_{\mu} \, \cos\theta \, S(x)}
     {D(x, \, \theta)}, 
     \label{eq:Ano03}
\end{eqnarray}
\noindent
where the denominator $D(x, \, \theta)$ is defined 
from Eq.~(\ref{eq:3no10}) as follows:
\begin{equation}
D(x, \, \theta) = 
     \frac{1}{x_{p}} \, {\cal D}_{0 \, 0 \, 0}(x, \, \theta) = 
     \frac{1}{x_{p}}
     [{\cal N}_{0 \, 0}(x) + P_{\mu} \, \cos\theta \,
     {\cal P}_{0 \, 0 \, 0}(x)].
      \label{eq:Ano04}
\end{equation}

The isotropic part $Q(x)$ and anisotropic part $S(x)$ 
of the longitudinal polarization are, respectively, expressed 
as follows:
\begin{eqnarray}
Q(x) &=& \xi^{(0 \, 0 \, 0)} \, \left[ q_{1}(x) +
     q_{2}(x) \, \delta^{(0)} \right],
     \label{eq:Ano05} \\
S(x) &=& \left[ s_{1}(x) + s_{2}(x) \, \rho^{(0 \, 0)} +
     s_{3}(x) \, \eta^{(0 \, 0)} \right].
     \label{eq:4no06}
\end{eqnarray}
where
\begin{eqnarray}
q_{1}(x) &=& x_{p} \, (3 - 2 \, x - r_{0}^{2}),
     \label{eq:Ano07} \\
q_{2}(x) &=& 12 \, x_{p} \, (1 - x),
     \label{eq:Ano08} \\
s_{1}(x) &=& ( - x + 2 \, x^{2} - x_{0}^{2} ),
     \label{eq:Ano09} \\
s_{2}(x) &=& 12 \, x (1 - x),
     \label{eq:Ano10} \\
s_{3}(x) &=& - 2  \, x_{0} \, (1 - x).
     \label{eq:Ano11}
\end{eqnarray}
The parameters ($\xi^{(0 \, 0 \, 0)}$ and $\delta^{(0)}$) in $Q(x)$ 
are defined in Eq.~(\ref{eq:2no23}) 
for the case of ${\cal P}(x)$, while the parameters ($\rho^{(0 \, 0)}$ 
and $\eta^{(0 \, 0)}$) in $S(x)$ are defined in Eq.~(\ref{eq:2no19}) 
for the case of ${\cal N}(x)$.

%%%%%%%%%%%%%%%%%%%% A-2 %%%%%%%%%%%%%%%%%%%%%%%%%%%%%%%%%%%%%%%
\subsection{Transverse polarization within the decay plane: 
$P_{T1}(x, \, \theta)$}
%%%%%%%%%%%%%%%%%%%% A-2 %%%%%%%%%%%%%%%%%%%%%%%%%%%%%%%%%%%%%%%

The $x$-dependent part $R(x)$ of $P_{T1}(x, \, \theta)$ is 
defined as
\begin{eqnarray}
P_{T1}(x, \, \theta)
  =  \frac{P_{\mu} \, \sin\theta \, R(x)}
     {D(x, \, \theta)}
     \label{eq:Ano12}
\end{eqnarray}
with
\begin{eqnarray}
R(x) = \left[ r_{1}(x) \, (1 - 12 \, \rho^{(0 \, 0)}) +
     r_{2}(x) \, \eta^{(0 \, 0)} \right],
     \label{eq:4no13}
\end{eqnarray}
where
\begin{eqnarray}
r_{1}(x) &=& - x_{0} \, (1 - x),
     \label{eq:Ano14} \\
r_{2}(x) &=& - 2 \, (x - x_{0}^{2}).
     \label{eq:Ano15}
\end{eqnarray}

Note that the small quantity $x_{0}$ appears in $r_{1}(x)$, 
which stands for the prediction from the standard model.  Meanwhile, 
$\eta^{(0 \, 0)}$ which indicates the existence of the Majorana 
neutrino is associated with the larger deviation function 
$r_{2}(x)$.

%%%%%%%%%%%%%%%%%%%% A-3 %%%%%%%%%%%%%%%%%%%%%%%%%%%%%%%%%%%%%%%
\subsection{Transverse polarization perpendicular to the 
decay plane: $P_{T2}(x,\theta)$}
%%%%%%%%%%%%%%%%%%%% A-31 %%%%%%%%%%%%%%%%%%%%%%%%%%%%%%%%%%%%%%%

The $x$-dependent part $T(x)$ of $P_{T2}(x, \, \theta)$ is 
defined as
\begin{eqnarray}
P_{T2}(x, \, \theta)
  = \frac{P_{\mu} \, \sin\theta \, T(x)}
    {D(x, \, \theta)},
     \label{eq:Ano16}
\end{eqnarray}
where
\begin{eqnarray}
T(x) = 2 \, x_{p} \left( 1 - r_{0}^{2} \right)
     \, \eta^{(0 \, 0)}_{im}.
     \label{eq:Ano17}
\end{eqnarray}
Here the new parameter $\eta^{(0 \, 0)}_{im}$ is defined as 
follows:~\cite{Doi2005}
\begin{eqnarray}
\eta^{(0 \, 0)}_{im}
     = \varepsilon_{m} \, 
     \left( \frac{\lambda}{{\cal A}^{(0 \, 0)}} \right)
     \mbox{Im} ( \overline{w_{e \mu}}^{\, *} \,\,
     \overline{w_{e \mu \, h}} ).
     \label{eq:Ano18}
\end{eqnarray}

A non-zero value of $T(x)$ implies 
the existence of a non-zero Majorana \textit{CP} violating 
phase in our model. There is no corresponding term in 
either the standard model or our model for the Dirac neutrino.

%%%%%%%%%%%%%%%%%%%%%%%%%%%%%%%%%%%%%%%%%%%%%%%%%%%%%%%%%%%%%%%
%%%%%%%%%%%%%%%%%         References         %%%%%%%%%%%%%%%%%%
%%%%%%%%%%%%%%%%%%%%%%%%%%%%%%%%%%%%%%%%%%%%%%%%%%%%%%%%%%%%%%%

%%% we added errata in  Sep.,2009
\vspace{1cm}
\begin{center}
------------  {\bf Errata added in September, 2009}  ------------\\
\end{center}
\vspace{0.4cm}
\noindent
1. \hspace{1mm} The inequality $\rho^{(s \, t)} > 0$ in 
Eq.~(2$\cdot$25) should be replaced by $\rho^{(s \, t)} \ge 0$.
\vspace{3mm}

\noindent
2. \hspace{1mm} Equation (2$\cdot$36) in \S 2 should be changed 
from $\eta^{(2 \, 0)}=(1 \pm 24)\cdot 10^{-3}$ to
$$
    \eta^{(2 \, 0)} = -0.12 \pm 0.21 \;\; \mbox{or} \;\;
     - (2.1 \pm 7.0  \pm 1.0) \cdot 10^{-3} 
     \hspace{3.5mm} \mbox{for the Majorana neutrino case.}
     \label{R20} \eqno(\mbox{E$\cdot$1})
$$

\noindent
Here, the former $\eta^{(2 \, 0)} = -0.12 \pm 0.21$ is 
reported by Derenzo$^{1)}$ from 
his analysis of the $e^{+}$ energy spectrum.  The theoretical expression 
he used is identical to our $N_{20}(x)$ in Eq.~(2$\cdot$24). 
On the other hand, the latter is obtained by Danneberg et 
al.$^{2)}$ through their restricted analysis of the 
transverse polarization $P_{T1}(x, \theta = \pi / 2)$ of $e^{+}$.  
Their approximated theoretical expression  
is the same as ours under the assumption $\rho^{(2 \, 0)}(x) = 0$, 
because they use the one-parameter fitting by 
assuming $\rho_M=0.75$ (i.e., $\rho^{(2 \, 0)} = 0$).  Our 
expression of $P_{T1}$ in the $(s, \, t, \, u) = (2, \, 0, \, 6)$ 
case is expressed without any approximation as
$$
    P_{T1}(x, \theta = \pi / 2) =
     \frac{P_{\mu} \, R^{(20)}(x)}
     {\; \; D^{(206)}(x, \theta = \pi/2) \;\;},
     \label{VA-PT1} \eqno(\mbox{E$\cdot$2})
$$
where
$$
    R^{(20)}(x) = - x_0 (1-x) (1 - 14 \rho^{(20)})
     - 2 (x-x_0^2) \eta^{(20)},
     \label{R} \eqno(\mbox{E$\cdot$3})
$$
$$
    D^{(206)}(x, \theta = \pi/2) =
     \left( 3x -2x^{2} - x_{0}^{2} \right)
     +2 \left( 3 x - 4 x^{2} + x_{0}^{2} \right) \rho^{(20)}
     + 6 x_{0} (1-x) \eta^{(20)}.
     \label{D} \eqno(\mbox{E$\cdot$4})
$$

\noindent
Note that our expression of $P_{T1}$ in the 
$(s, \, t, \, u) = (0, \, 0, \, 0)$ case 
is given in Eq.~(A$\cdot$12) in Appendix A.

The data $\eta^{(2 \, 0)}=(1 \pm 24)\cdot 10^{-3}$ that we  
cited in our paper is that reported by the Particle Data 
Group$^{3)}$ as an average over the values of $\eta$ 
obtained by different 
experiments.  This citation is not appropriate, because some of 
them are derived from the data on $P_{T1}(x, \theta)$ by 
using theoretical expressions different from our $R^{(20)}(x)$ 
in Eq.~(E$\cdot$3).\\
\begin{center}
{\bf Acknowledgements}  \\
\end{center}
%Acknowledgements:\\

We would like to thank Prof. A. Olin for pointing out our inadequate 
citation of the experimental data for $\eta^{(2 \, 0)}$.
%\section{Second Appendix}
\begin{center}
{\bf References}  \\
\end{center}
1)\ 
S.~E.~Derenzo,~Phys.~Rev. \ {\bf 181}~(1969),~1854.\\
2)\ 
N.~Danneberg et al.,~Phys.~Rev.~Lett. \ {\bf 94}~(2005),~021802.\\
3)\ 
For general review, see Particle Data Group (Review: W.~Fetscher 
and H.~J.~Gerber), J.~Phys.\,G:~Nucl.~Part.~Phys. {\bf 33}~(2006),~440.\\
\end{document}